\documentclass[11pt,a4paper]{article}
\pdfoutput=1

\usepackage{jcappub}
\usepackage{amsfonts,amssymb,amsmath,graphicx,color,bm,comment,soul}
\definecolor{ultramarine}{rgb}{0.07, 0.04, 0.56}
\definecolor{cadmiumgreen}{rgb}{0.0, 0.42, 0.24}
\definecolor{indigo(dye)}{rgb}{0.0, 0.25, 0.42}
\definecolor{orangered}{RGB}{255,69,0}
\usepackage[linktocpage=true]{hyperref}
\hypersetup{
colorlinks=true,
citecolor=ultramarine,
linkcolor=cadmiumgreen,
urlcolor=indigo(dye),
}

\newcommand{\f}[2]{\frac{#1}{#2}}  
\newcommand{\mk}[1]{\left( #1 \right)}  
\newcommand{\kk}[1]{\left[ #1 \right]}  
  
\newcommand{\be}{\begin{equation}}  
\newcommand{\ee}{\end{equation}}
\newcommand{\Mpl}{M_{\rm Pl}}
\newcommand{\pa}{\partial}
\newcommand{\A}{\mathcal{A}}
\newcommand{\diff}[2]{\ensuremath{\frac{\text{d}#1}{\text{d}#2}}}
\newcommand{\difs}[2]{\ensuremath{\frac{\text{d}^2 #1}{\text{d}#2^2}}}
\newcommand{\dif}{\text{d}}
\newcommand{\logn}{\text{ln\,}}

\notoc



\begin{document}

\subheader{YITP-18-26}

\title{Reconciling tensor and scalar observables in G-inflation}

\author[a,b]{H\'ector Ram\'irez}
\author[c]{, Samuel Passaglia}
\author[d]{, Hayato Motohashi}
\author[c]{, Wayne Hu}
\author[b]{and Olga Mena}

\affiliation[a]{Departamento de F\'{i}sica Te\'orica, Universidad de Valencia,\\ 
Dr. Moliner 50, E-46100 Burjassot, Spain}
\affiliation[b]{Instituto de F\'{i}sica Corpuscular (IFIC), Universidad de Valencia-CSIC,\\
Catedr\'atico Jos\'e Beltr\'an 2, E-46980, Paterna, Spain}
\affiliation[c]{Kavli Institute for Cosmological Physics \textit{\&} Department of Astronomy and Astrophysics,\\ The University of Chicago,\\ 
5640 South Ellis Avenue, Chicago, Illinois 60637, USA}
\affiliation[d]{Center for Gravitational Physics, Yukawa Institute for Theoretical Physics,\\
Kyoto University,\\ 
Kitashirakawa-Oiwakecho, Sakyo-Ku, Kyoto 606-8502, Japan}

\emailAdd{hector.ramirez@uv.es}
\emailAdd{passaglia@uchicago.edu}
\emailAdd{hayato.motohashi@yukawa.kyoto-u.ac.jp}
\emailAdd{whu@background.uchicago.edu}
\emailAdd{omena@ific.uv.es}

\abstract{The simple $m^2\phi^2$ potential as an inflationary model is coming under increasing tension with limits on the tensor-to-scalar ratio $r$ and measurements of the scalar spectral index $n_s$. Cubic Galileon interactions in the context of the Horndeski action can potentially reconcile the observables. However, we show that this cannot be achieved with only a constant Galileon mass scale because the interactions turn off too slowly, leading also to gradient instabilities after inflation ends. Allowing for a more rapid transition can reconcile the observables but moderately breaks the slow-roll approximation leading to a relatively large and negative running of the tilt $\alpha_s$ that can be of order $n_s-1$. We show that the observables on CMB and large scale structure scales can be predicted accurately using the optimized slow-roll approach instead of the traditional slow-roll expansion. Upper limits on $|\alpha_s|$ place a lower bound of  $r\gtrsim 0.005$ and, conversely, a given $r$ places a lower bound on $|\alpha_s|$, both of which are potentially observable with next generation CMB and  large scale structure surveys.}

\keywords{inflation, cosmological parameters from CMBR}

\arxivnumber{1802.04290}

\maketitle

\section{Introduction}
\label{sec:int}

Inflation is a leading paradigm able to solve the main problems of the standard model of cosmology and, at the same time, able to generate the quantum seeds that could have given rise to the structures we see today in the sky. The canonical picture consists of introducing a new scalar field minimally coupled to Einstein gravity, the inflaton, which drives the expansion of the Universe from quantum to cosmological scales at an exponential rate while it slowly rolls towards the minimum of its potential. This potential is required to be sufficiently flat in order to have enough time to form a Universe consistent with the isotropy and homogeneity observed today. Although the paradigm itself is consistent with the latest observational constraints on the scalar and tensor power spectra (see {\it e.g.}~\cite{Ade:2015lrj}), simple quadratic and monomial potentials are coming into increasing conflict with these constraints.

Inflationary models with noncanonical terms can arise naturally from particle physics and allow more freedom to
satisfy observational constraints~\cite{Alishahiha:2004eh, Silverstein:2003hf, Carrasco:2015pla, Bean:2008na, Silverstein:2008sg, McAllister:2008hb, Flauger:2009ab}. Models with nonminimal couplings, for instance, are able to reconcile with current measurements some of the earliest and simplest realizations of inflation, such as those with power-law potentials~\cite{Futamase:1987ua, Fakir:1990eg, Komatsu:1999mt, Hertzberg:2010dc, Okada:2010jf, Linde:2011nh, Boubekeur:2015xza}.

General scalar-tensor theories of gravity provide a unified framework upon which one can construct new models of inflation or embed known ones in a broader context. The most general four-dimensional scalar-tensor theory in curved space-time which leads to second-order equations of motion -- thus free from ghosts and related instabilities -- is the Horndeski~\cite{Horndeski:1974wa}, or generalized Galileon~\cite{Nicolis:2008in,Deffayet:2011gz,Kobayashi:2011nu}, theory\footnote{While healthy theories beyond Horndeski have been developed to include higher derivatives in the equations of motion (see \cite{Zumalacarregui:2013pma,Gleyzes:2014dya,Motohashi:2014opa,Langlois:2015cwa,Motohashi:2016ftl,Klein:2016aiq,BenAchour:2016fzp,Crisostomi:2017aim,Motohashi:2017eya}), we will restrict our analysis to models within the Horndeski framework.}. Recently there have been efforts to construct models of so-called G-inflation using the Horndeski Lagrangian by explicitly choosing the form of the independent functions of the scalar field and its derivatives.  Such models must be carefully constructed to avoid instabilities, given that the Galilean symmetry should be broken in order to have a successful inflation~\cite{Kobayashi:2010cm,Burrage:2010cu, Kamada:2010qe, Kobayashi:2011nu, Ohashi:2012wf, Kamada:2013bia}.

When constructing phenomenologically viable models in  the more general parameter space,  the usual slow-roll approximation may not always suffice to describe observables.   While numerically solving the scalar and tensor equations of motion is always possible, generalized slow-roll (GSR) techniques have been developed to overcome the deficiencies of the traditional slow-roll approach~\cite{Stewart:2001cd, Kadota:2005hv, Dvorkin:2009ne, Hu:2011vr, Hu:2014hoa, Motohashi:2015hpa, Miranda:2015cea}. In particular the optimized slow-roll (OSR) expansion of GSR provides an improved way of evaluating scalar and tensor spectra for inflationary models with slow-roll violation on a time scale of a few $e$-folds or larger~\cite{Motohashi:2015hpa}. Recently these approaches have been extended to cover the full space of Horndeski models, allowing one to compute the inflationary observables without imposing the slow-roll conditions~\cite{Motohashi:2017gqb}. Their efficacy have been tested for large slow-roll violations such as those required by primordial black hole (PBH) formation models~\cite{Motohashi:2017kbs}.

In this paper we show that it is possible to reconcile the observational tension between scalar and tensor observables in $m^2\phi^2$ inflation by introducing a transient G-inflation regime, for which the GSR and OSR formulas provide a good description of inflationary observables. In \S\ref{sec:mod} we review the Horndeski Lagrangian and show why  simple models with a constant Galileon interaction mass scale introduced in previous studies \cite{Ohashi:2012wf} can no longer satisfy the latest observational constraints. In \S\ref{sec:step} we show how to overcome these difficulties by introducing a transition during inflation that transiently violates the
slow-roll approximation. In \S\ref{sec:gsr} we show how the {GSR and OSR} techniques accurately relate the parameters of these models to the scalar and tensor observables. Finally, we conclude in \S\ref{sec:con}.

\section{Potential-driven G-Inflation}
\label{sec:mod}

Horndeski gravity is the most general scalar-tensor theory in four dimensions which leads to second-order equations of motion. The full Lagrangian is given by
\begin{align} \label{eq:Horn}
\mathcal{L_H} & = G_2+ G_3\Box\phi+ G_4R \notag\\
&~~~- 2 G_{4,X} [ (\Box\phi)^2 -\phi^{;\mu\nu}\phi_{;\mu\nu}] + G_5 G^{\mu\nu} \phi_{;\mu\nu} \\ 
&~~~+ \f{G_{5,X}}{3} [ (\Box\phi)^3 - 3 (\Box\phi) \phi_{;\mu\nu}\phi^{;\mu\nu} + 2 \phi_{;\mu\nu}\phi^{;\mu\sigma} {\phi^{;\nu}}_{\!;\sigma}]~, \notag
\end{align} 
where $G_n=G_n(\phi,X)$ are arbitrary functions of $\phi$ and $X\equiv g^{\mu\nu} \pa_\mu\phi\pa_\nu\phi$, $G_{n,X}\equiv \pa G_n/\pa X$, $\phi_{;\mu\nu}=\nabla_\mu\nabla_\nu\phi$ and $R$ and $G_{\mu\nu}$ are the Ricci and Einstein tensors respectively. For $G_2=-X/2-V(\phi)$, $G_4=1/2$, and $G_3=G_5=0$, we recover the Lagrangian for canonical inflation\footnote{Here and throughout we take units where $\Mpl=1/\sqrt{8\pi G}=1$.}.

From Eq.~(\ref{eq:Horn}) one can now choose the $G_n$ functions to construct more general phenomenological models of inflation given that the simplest realizations are being ruled out by the latest cosmological measurements. For instance,  the chaotic inflation model provides a large value for the tensor-to-scalar ratio $r$ which is disfavored by current observations. Ref.~\cite{Ohashi:2012wf} showed that with the introduction of a $G_n$ term the relationship between the tensor and scalar observables can be modified.  However, we shall now see that under the slow-roll approximation, this additional freedom is not sufficient to reconcile observations with the predictions of chaotic inflation.

Concretely, Ref.~\cite{Ohashi:2012wf} considered a model of potential-driven G-inflation of the form
\begin{align} \label{eq:gns}
G_2(\phi,X) &= -\f{X}{2} - V(\phi)~, \notag\\ 
G_3(\phi,X) &=f_3 \f{X}{2}~, \notag\\
G_4(\phi,X) &= \f{1}{2}~,\\ 
G_5(\phi,X) &= 0~,\notag
\end{align}
with a chaotic inflation potential $V(\phi)=m^2\phi^2/2$ and $f_3=-M^{-3}$, where $m$ and $M$ are the inflaton and Galileon mass scales respectively\footnote{G-inflation was originally introduced in ~\cite{Kobayashi:2010cm,Kobayashi:2011nu} as a model for inflation driven kinetically by the Galileon field. The models discussed here, on the other hand, are potential-driven versions, first studied in \cite{Kamada:2010qe}.}.

Taking Eqs.~(\ref{eq:gns}), assuming the general case in which $f_3=f_3(\phi)$, and working on the flat Friedmann-Lema\^itre-Robertson-Walker (FLRW) metric,
\be \dif s^2= -\dif t^2 + a(t)^2 \delta_{ij} \dif x^i \dif x^j, \ee
the Einstein and Klein-Gordon equations can be written as
\begin{align} \label{eq:bground}
V & -\mk{3-\frac12\phi'^2}H^2+\mk{3f_3\phi'^3-\frac{f_{3,\phi}}{2}\phi'^4}H^4=0~, \notag\\ 
V & -2H' H-\mk{3+\frac12\phi'^2}H^2+ f_3H^3 H'\phi'^3 +\mk{f_3\phi'^2\phi''+\frac{f_{3,\phi}}{2}\phi'^4}H^4=0~,\notag\\ 
V &_{,\phi} + HH'\phi'+\mk{3\phi'+\phi''}H^2+\mk{9f_3H'\phi'^2-2f_{3,\phi}H'\phi'^3}H^3 \\ 
&+\mk{9f_3\phi'^2+6f_3\phi'\phi''-2f_{3,\phi}\phi'^2\phi''-\frac{f_{3,\phi\phi}}{2}\phi'^4}H^4=0~,\notag
\end{align}
where $H$ is the Hubble parameter and derivatives are defined as $'\equiv\dif{}/\dif{N}$, being $\dif{N}\equiv H\dif{t}=(H/\dot\phi)\dif{\phi}$ the number of $e$-foldings of inflation, and $_{,\phi}\equiv\dif{}/\dif{\phi}$.

In the slow-roll (SR) approach, Eqs.~(\ref{eq:bground}) may be approximated as \cite{Ohashi:2012wf}
\begin{align}
\label{eq:slowroll}
&3H^2 \approx V,\notag\\
&3H^2\phi'(1+\A)+V_{,\phi} \approx 0~,
\end{align}
and
\be \label{eq:epsilonSR}
\epsilon_H \equiv
 -\frac {H'}{H} \approx \frac{1}{2(1+\A)} \left( \frac{V_{,\phi}}{V} \right)^2~.
\ee
Here 
\be \label{Adef} \A\equiv 3f_3H^2\phi' \ee 
measures the deviation from  canonical inflation: for $|\A| \ll 1$ the Galileon term produces negligible effects. In \S\ref{ssec:gsbg}, 
we use this slow-roll approximation for $\epsilon_H$, Eq.~\eqref{eq:epsilonSR}, as a test of the slow-roll approximation itself. For the chaotic inflation potential, $\phi' <0$ and thus if $f_3 <0$ the combination of Eqs.~\eqref{eq:slowroll} and \eqref{Adef} gives 
\begin{align}
\A \approx \frac{\sqrt{1- 4 f_3 V_{,\phi}}-1}{2}.
\label{eq:Aslowroll}
\end{align}

The original G-inflation model, hereafter called the ``G-model," took a constant $f_3=-M^{-3}$ so that far up the potential or at early times the Galileon term dominates, whereas the canonical terms come to dominate as the field rolls down. The transition between the two regimes is marked by $\A=1$ where $V_{,\phi} = -2/f_3 = 2 M^3$ ~\cite{Ohashi:2012wf}. It is therefore interesting to consider the relationship between the tensor and scalar observables as a function of $\A$. The scalar and tensor power spectra, under the slow-roll approximation, can be written as \cite{Ohashi:2012wf}
\begin{align} \label{eq:srpsfor}
\Delta_\zeta^{2\,\text{(SR)}}&=\f{V^3}{12\pi^2V_{,\phi}^2}\f{\mk{1+\A}^2\mk{1+2\A}^{1/2}}{\mk{1+4\A/3}^{3/2}}~,\notag\\
\Delta_\gamma^{2\,\text{(SR)}}&=\f{V}{6\pi^2}~,
\end{align}
where the tensor power spectrum is defined for each polarization state separately and is not modified from its form in canonical inflation.  Therefore for the same position on the potential in field space, the G-model enhances scalar power over tensor power linearly in $\A$ for $\A \gg 1$.

However, given the strong experimental constraints on the tilt, the tensor-to-scalar ratio of the G-model should be compared to chaotic inflation at the same tilt rather than the same field value. The scalar tilt and tensor-to-scalar ratio are defined as usual as
\begin{align}
n_s-1  &\equiv \frac{\dif\ln \Delta^2_{\zeta}}{\dif\ln k}~,\\
r& \equiv 4\,\f{\Delta_\gamma^2}{\Delta_\zeta^2}~.
\label{eq:rns}
\end{align}
For comparison to the CMB observables, these should be evaluated at $k=k_*=0.05$ Mpc$^{-1}$. These evaluations require converting a given field value $\phi$ to a wavenumber $k$. Under slow-roll, scalar fluctuations freeze out\footnote{Note that the scalar sound speed is $c_s=\sqrt{2/3}$ for $\A \gg 1$ and $c_s=1$ for $\A\ll 1$ whereas the tensor sound speed is $c_t=1$. Even in slow-roll, the freeze-out condition should in principle differ between the two as we discuss below, but given slow variation of the expressions in \eqref{eq:Aslowroll} and \eqref{eq:srpsfor}, Ref.~\cite{Ohashi:2012wf} ignored these distinctions.}   when $c_s k=aH$, and therefore this relationship requires a mapping between field values and the number of $e$-folds to the end of inflation  $\Delta N= N_f-N$.  From Eqs.~\eqref{eq:bground} and \eqref{Adef},
\be
\label{eq:DN}
\Delta N\simeq\int^\phi_{\phi_f}(1+\A)\f{V}{V_{,\tilde{\phi}}}\dif{\tilde{\phi}}~.
\ee
Putting these relations together
Ref.~\cite{Ohashi:2012wf} found for $\A \gg 1$, 
\be
n_s-1=-\f{9}{5\Delta N+2}~,\qquad r=\f{64\sqrt{6}}{9}\f{2}{5\Delta N+2}~\notag,
\ee
where, by eliminating the $e$-folds to the end of inflation, we obtain the parametric relation
\be
r=-\f{128}{27}\sqrt{\f{2}{3}}(n_s-1)~.
\label{eq:rnslimG}
\ee
For the $\A \ll 1$ limit, one recovers the canonical chaotic predictions
\be
n_s-1=-\f{4}{2\Delta N+1}~,\qquad r=\f{16}{2\Delta N+1}~\notag,
\ee
which combined give
\be
r=-4(n_s-1) \,.
\label{eq:rnslimC}
\ee
Ref.~\cite{Ohashi:2012wf} noted that for a fixed $e$-fold, $\Delta N\sim 50-60$, the $\A \gg 1$ case has a lower $r$ and larger $n_s$. However we see from Eqs.\ \eqref{eq:rnslimG} and \eqref{eq:rnslimC} that for the same $n_s$, the $\A\gg 1$ limit lowers $r$ only by a negligible factor of $\approx 0.97$. With recent improvements in the constraints on both parameters, the G-model cannot cure the $r$-$n_s$ problem of the canonical $\phi^2$ model given any choice of $M$ or $\Delta N$. Furthermore no smooth transition or interpolation between these two very close forms can solve this problem either. Monomial potentials with steeper indices than $\phi^2$ face a similar issue.

While this might seem like a no-go for simple $\phi^n$ potentials, we will show in the following sections that a more rapid transition between these two limits provides a solution where the scalar tilt is substantially but transiently lowered while $\A$ remains sufficiently large to suppress $r$. Furthermore by allowing a more rapid transition, we automatically cure the gradient instability problem for these models.  This problem arises if the transition to $\A<1$ occurs after the end of inflation such that the scalar sound speed squared $c_s^2$ oscillates and becomes negative during reheating. In the original G-inflation model, this restriction places a lower limit on $M$ \cite{Ohashi:2012wf} and an upper limit on the enhancements to the scalar power spectrum through $\A$.
However, by making the transition more rapid, we can make it complete before the end of inflation for any $M$\footnote{In \cite{Kamada:2013bia}, the addition of a kinetic $X^2$ term to the Lagrangian was proposed. This term adds a positive contribution to $c_s^2$, thus removing gradient instabilities. However, the effect of the Galileon term was weakened.}.

\section{Potential-driven G-Inflation with step}
\label{sec:step}

As discussed in the previous section, the phenomenological problems of the original version of G-inflation 
arise because the transition to canonical inflation takes place too slowly.  To resolve these problems, we promote $f_3$ in Eq.~\eqref{eq:gns} to be a step-like function of $\phi$, hereafter called the ``Step model",
\be
f_3(\phi)=-M^{-3}\kk{1+\text{tanh}\mk{\f{\phi-\phi_r}{d}}}~,
\label{eq:step}
\ee
where $\phi_r$ and $d$ are new parameters of the model related to the position in field space and the width of the step respectively. This allows us to control the epoch and the rapidity of the transition from G-inflation to canonical inflation. By making this transition sufficiently rapid we can evade the observational problems in the $r$-$n_s$ plane as well as eliminate the gradient instabilities at the end of inflation.

\begin{figure}[t]
\centering
\includegraphics[keepaspectratio, width=12cm]{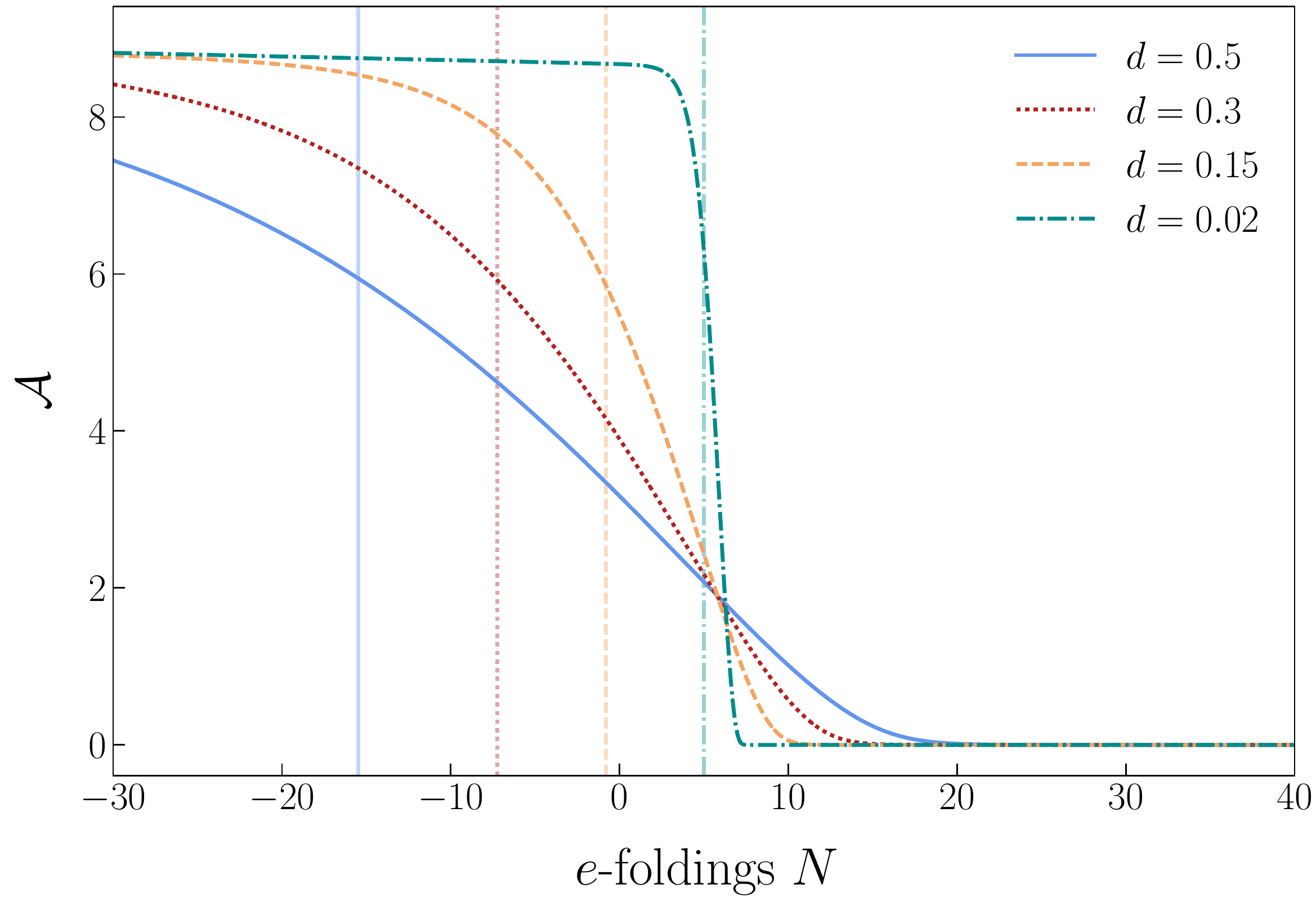}
\caption{G-inflation transition parameter $\A$ from Eq.~(\ref{Adef}) as a function of $e$-folds $N$, for the model given in Eq.~(\ref{eq:step}) and for the values $M=1.303\times10^{-4}$, $m=2.58\times10^{-6}$, $\phi_r=13.87$ and four different values of the step width $d$: $d=\{0.5,0.3,0.15,0.02\}$.  Vertical lines denote $N(\phi_r)$, the epoch at which the inflaton crosses the center of the step.}
\label{fig:A}
\end{figure}

\subsection{Background transition}
\label{ssec:gsbg}

With $f_3(\phi)$ given in Eq.~\eqref{eq:step}, we can numerically solve the background equations \eqref{eq:bground} following the procedure explained in \cite{Ohashi:2012wf}. As discussed in \S\ref{sec:mod}, the transition from G-inflation to canonical inflation is controlled by $\A$ in Eq.~\eqref{Adef}: namely, $\A$ evolves from $\A\gg 1$ to $\A\ll 1$, with the transition occuring at  $\A \approx 1$.   For the model in Eq.~\eqref{eq:step}, the rapidity of the transition is controlled by the step width $d$. Figure~\ref{fig:A} shows the evolution of $\A$  for different values of $d$ with $m$, $M$ and $\phi_r$ fixed to values which we will motivate below. One can see that the transition takes fewer $e$-folds $N$ for a sharp step, {\it i.e.} for a small $d$. In these Step model examples $N$ is defined in such a way that at the end of inflation $N_f=55$. We then take $N=0$ as the epoch when CMB scales or specifically  $k_*=0.05$ Mpc$^{-1}$ left the scalar sound horizon
\begin{equation} \label{eq:horizon}
\int_{0}^{55} \dif{N} \, \frac{c_s}{aH}=20\, {\rm Mpc}~.
\end{equation}
Note that the wider the step is, the more the $e$-fold for which $\A = 1$  lags $N(\phi_r)$ (shown with vertical lines), when the inflaton passes the center of the step.

\begin{figure}
\centering
\includegraphics[keepaspectratio, width=12cm]{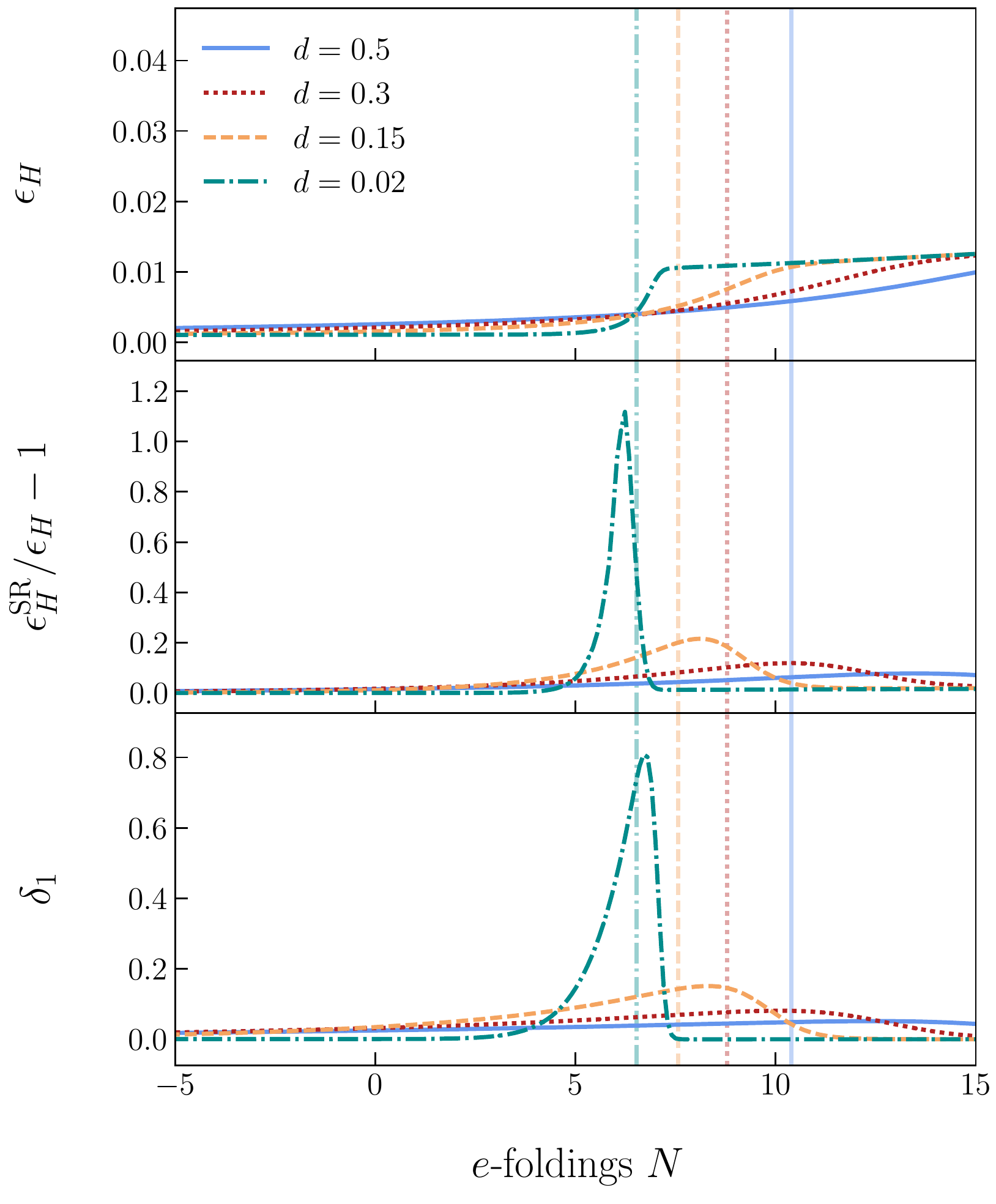}
\caption{Exact solution for the slow-roll prediction of $\epsilon_H$ (RHS of Eq.~\eqref{eq:epsilonSR}) (upper panel), fractional difference between the solution employing the approximation in Eq.~\eqref{eq:DN} and the exact background value (middle panel), and exact solution for the slow-roll parameter $\delta_1=\f{1}{2}\f{\textrm{d}\ln\epsilon_H}{\textrm{d}N}-\epsilon_H$ (lower panel). All as a function of $N$ and for the same models of Fig.~\ref{fig:A}. The vertical lines represent $N(\A=1)$ for each curve where the SR violation is nearly maximal.}
\label{fig:dN}
\end{figure}

With a rapid transition, we generically expect that the SR approximation will break down. In Fig.~\ref{fig:dN} we show the evolution of $\epsilon_H$ for the same cases as Fig.~\ref{fig:A} calculated numerically and through
the slow-roll approximation of  Eq.~\eqref{eq:epsilonSR}.  In the slow-roll comparisons here and below we use the numerical computation of $\phi(N)$ to avoid conflating errors in the mapping of Eq.~\eqref{eq:DN} and local deviations from slow-roll at a given $N$. Before and after the transition (but before the end of inflation), the slow-roll approximation is quite good. Near the transition, however, fractional differences increase as  $d$ decreases (Fig.~\ref{fig:dN}, middle panel). Furthermore, for a rapid evolution of $\epsilon_H$, it is expected that the second SR parameter, $\delta_1=\f{1}{2}\f{\textrm{d}\ln\epsilon_H}{\textrm{d}N}-\epsilon_H$, be of order $\sim1$, reaffirming the SR breakdown, as shown in the lower panel of Fig.~\ref{fig:dN}. In both cases, the SR deviations peak near the epoch when $\A=1$ (vertical lines). The rapid evolution of $\epsilon_H$ and corresponding breakdown of the slow-roll approximation requires going beyond the slow-roll approximation for the accurate calculation of scalar and tensor observables as we shall see in the next section.

\begin{figure}[t]
\centering
\includegraphics[keepaspectratio, width=12cm]{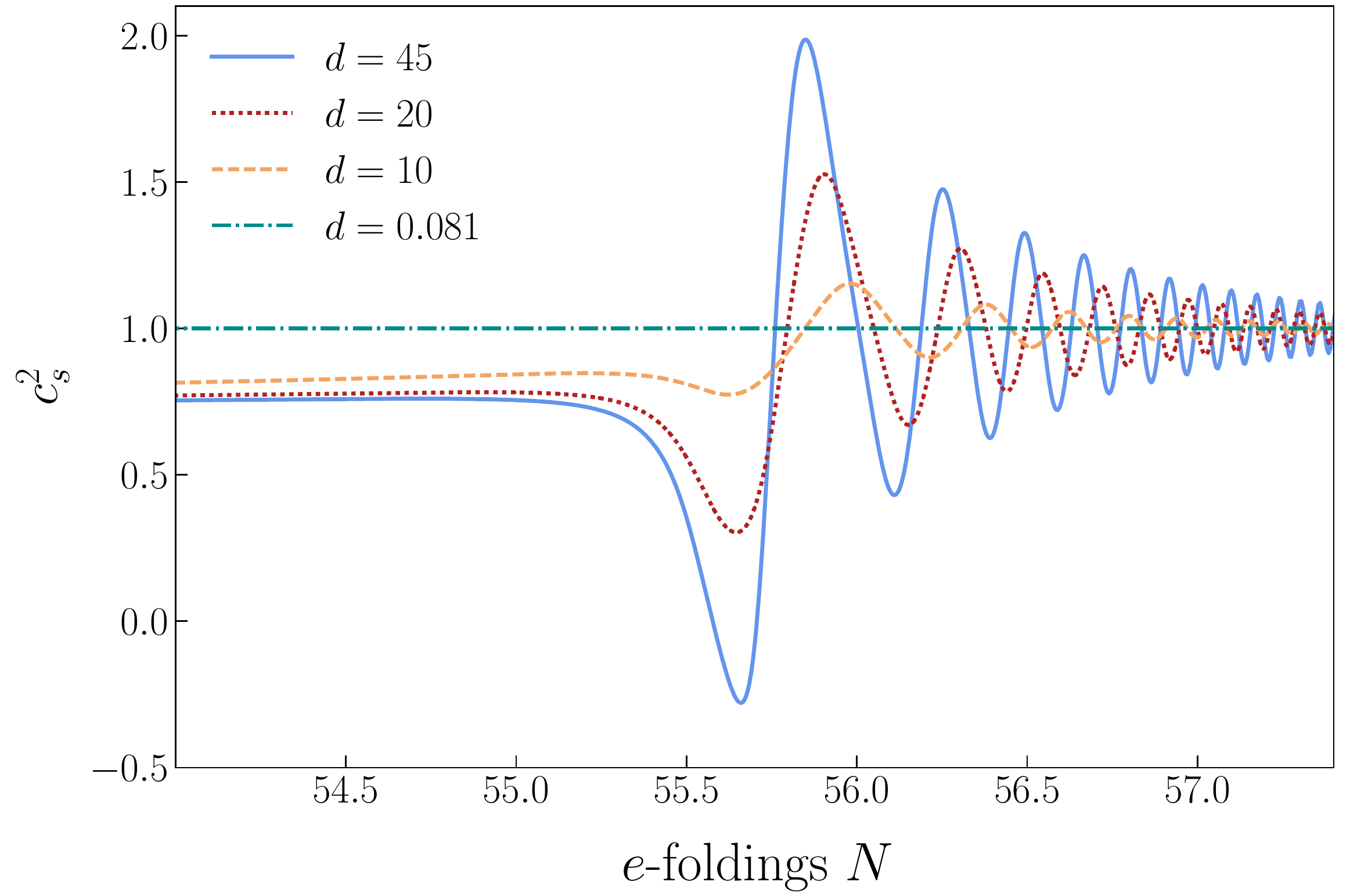}
\caption{Scalar sound speed squared near the end of inflation $N_f=55$
 for the choice of values $M=1.303\times10^{-4}$, $m=2.58\times10^{-4}$, $\phi_r=13.87$ and four different values of the step parameter $d$: $d=\{0.081, 10, 20, 45\}$.  Except for the widest case of $d=45$, the step ensures that the transition completes before the end of
 inflation,  $\A(\phi_f)<1$,  and eliminates the gradient instability, {\it i.e.}  $c_s^2>0$.}
\label{fig:cs}
\end{figure}

To finish the discussion on the background solutions, Fig.~\ref{fig:cs} shows the evolution of the sound speed squared of scalar perturbations, $c_s^2$, as a function of $N$ (see Eq.~\eqref{eq:cbs} for details). The value $M=1.303\times10^{-4}$ of the Galileon mass scale used here is below the lower limit obtained in~\cite{Ohashi:2012wf} corresponding to the avoidance of gradient instabilities in the G-model case. However, as expected for the Step model, we see that as long as the width is not very large that the transition fails to complete by the end of inflation, the gradient instabilities disappear -- $c_s^2$ is always positive -- and this holds independently of the value of the Galileon mass scale $M$. Since inflation ends at $\phi \sim 1$, this condition corresponds to setting the transition $\phi_r$ sufficiently large given the width $d$.  

\subsection{Inflationary observables}
\label{sec:obs}

In order to compute inflationary observables, we expand to quadratic order in scalar and tensor perturbations the action for $\mathcal{L_\mathcal{H}}$ given in Eq.~(\ref{eq:Horn}):
\begin{align} \label{eq:act}
S_\zeta^{(2)}&=\int\dif{}^4x\frac{a^3b_s\epsilon_H}{c_s^2}\mk{\dot\zeta^2-\frac{c_s^2k^2}{a^2}\zeta^2}~,\notag\\
S_\gamma^{(2)}&=\sum_{\lambda=+,\times}\int\dif{}^4x\frac{a^3b_t}{4c_t^2}\mk{\dot\gamma_\lambda^2-\frac{c_t^2k^2}{a^2}\gamma_\lambda^2}~,
\end{align}
parametrized by the sound speeds $c_{s,t}$ and normalization factors $b_{s,t}$ for scalar and tensor perturbations which contain all the information coming from the Horndeski framework~\cite{Motohashi:2017gqb}. For the choice given in Eqs.~(\ref{eq:gns}) these parameters can be computed as\footnote{For the general case in which none of the $G_n$ functions are taken to be equal to zero see \cite{Kobayashi:2011nu,Ohashi:2012wf}.}
\begin{align}\label{eq:cbs}
b_s&=\f{2 \mu_1H-2 \mu'_1H- \mu_1^2}{\epsilon_H }~,\notag\\ 
c_s^2&=\f{3\mk{2 \mu_1H-2 \mu'_1H-\mu_1^2}}{4 \mu_2+9\mu_1^2}~,\\
b_t&=1~,\qquad c_t^2=1~\notag,
\end{align}
where
\begin{align}
\mu_1&=2H-f_3H^3\phi'^3~, \notag\\
\mu_2&=-9H^2-3f_{3,\phi}H^4\phi'^4 +\f{3}{2}\mk{1+12H^2f_3\phi'}H^2\phi'^2~.
\end{align}
Notice that, for the choice of Eqs.~(\ref{eq:gns}), the tensor action is not modified from that of canonical inflation.

Varying the quadratic actions given in Eqs.~(\ref{eq:act}) we arrive at the Mukhanov-Sasaki equation
\be \label{eq:muksa}
\difs{u_i}{\tau}+\mk{c_i^2k^2-\f{1}{z_i}\difs{z_i}{\tau}}u_i=0~.
\ee
Here and below $\tau$ is the (positive, decreasing) conformal time until the end of inflation, and $i=s,t$ for the scalar and tensor perturbations
respectively.  We define the  Mukhanov-Sasaki variable as $u_s \equiv z_s\zeta$ and
$u_t\equiv z_t\gamma$ with 
\be
z_s=a\f{\sqrt{2b_s\epsilon_H}}{c_s}~,\qquad
z_t=\f{a}{c_t}\sqrt{\f{b_t}{2}}~,
\ee
for the quadratic actions in Eq.~\eqref{eq:act}.

As shown in Fig.~\ref{fig:dN}, for the Step model with a small step width we cannot apply the slow-roll approximation to solve Eq.~(\ref{eq:muksa}) due to the fact that the slow-roll conditions are violated near the transition where $\A \sim 1$. We instead solve this equation numerically from Bunch-Davies initial conditions to compute the power spectra as
\begin{align} \label{eq:standardps}
\Delta^2_{\zeta}(k)&=\lim_{{k\tau}\to0}\f{k^3}{2\pi^2}\big{\rvert}\zeta\big{\rvert}^2~,\notag\\
\Delta^2_{\gamma}(k)&=\lim_{{k\tau}\to0}\f{k^3}{2\pi^2}\big{\rvert}\gamma_{+,\times}\big{\rvert}^2~,
\end{align}
which define the inflationary parameters $n_s(k)$ and $r(k)$ through Eq.~\eqref{eq:rns}.

We now construct a working example of transient G-inflation in order to examine its observable phenomenology further. With the convention that the CMB mode exits the scalar sound horizon $55$ $e$-folds before the end of inflation, the Step model has four remaining free parameters: the mass scales $M$ and $m$, and the step parameters $\phi_r$ and $d$. The inflaton mass scale $m$ mainly controls the Hubble rate and hence the amplitude of the power spectra. We choose it to satisfy the {\it Planck} 2015 TT+lowP measurement of the scalar amplitude $A_s = \Delta^2_{\zeta}(k_*)= (2.198\pm0.08)\times10^{-9}$~\cite{Ade:2015lrj}. For a fixed $m$, the Galileon mass scale $M$ determines $\A$ when the CMB mode leaves the horizon, which sets the tensor amplitude relative to the scalar amplitude. We therefore fix it according to the desired suppression of $r$, for example $\A(0) \approx {8}$. Finally, the step parameters $\phi_r$ and $d$ are determined by the {\it Planck} 2015 TT+lowP scalar tilt $n_s = 0.9655\pm0.0062$ and bounds on the running of the tilt $\alpha_s=-0.0084\pm0.0082$. With four constraints for four parameters, we use slow-roll expressions to find initial parameter guesses which satisfy these conditions and then iterate using numerical results for the background and power spectrum (see \S\ref{sec:gsr}) to enforce the Planck constraints beyond slow-roll.

\begin{figure}[t]
\centering
\includegraphics[keepaspectratio, width=12cm]{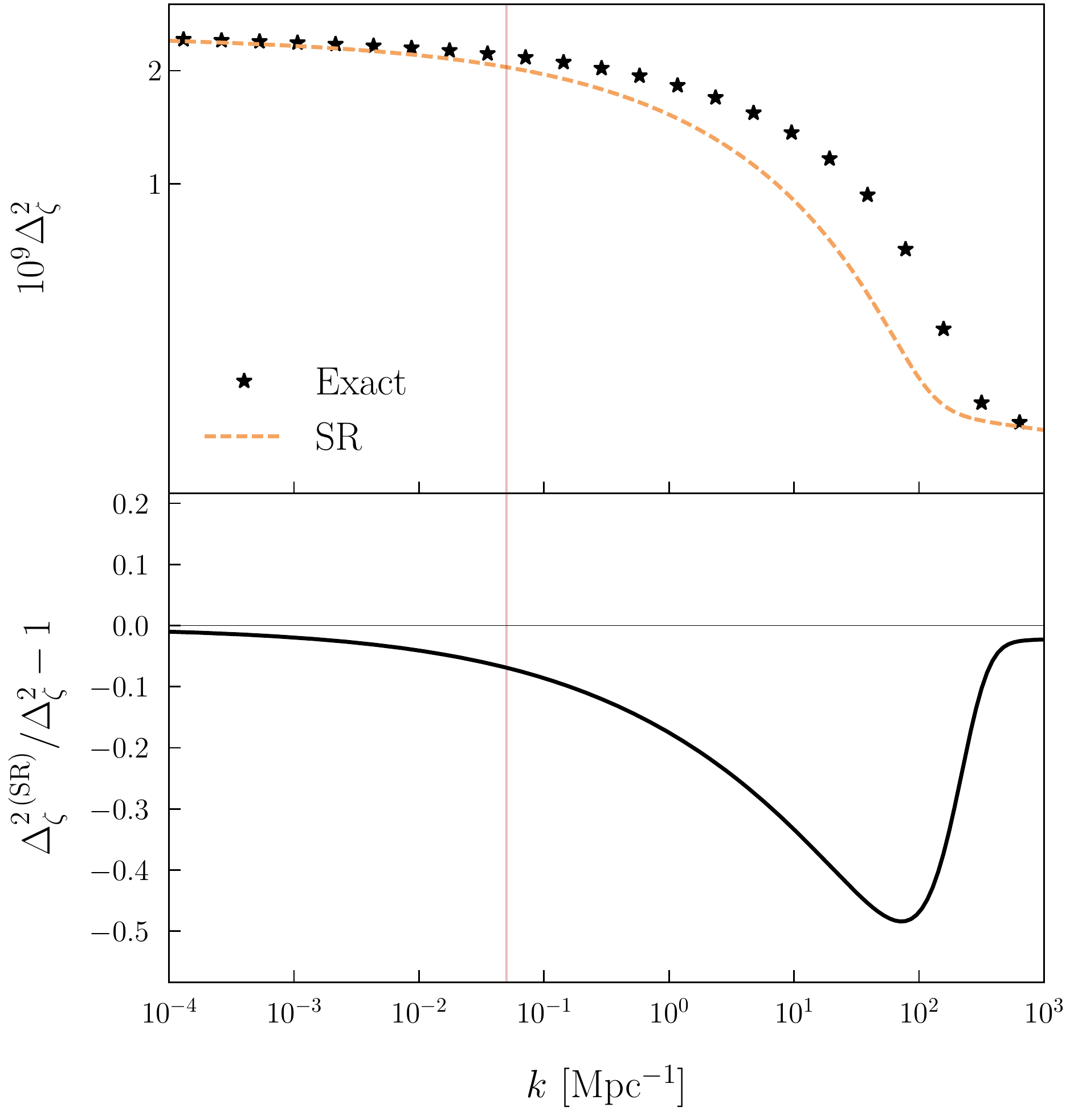}
\caption{Scalar power spectrum for the Step model computed by solving numerically Eq.~\eqref{eq:muksa} (stars) and by computing Eqs.~\eqref{eq:srpsfor} with the exact background solutions (dashed orange) (upper panel). Fractional difference between the two solutions (lower panel). The set of parameters used here is $M=1.303\times10^{-4}$, $m=2.58\times10^{-6}$, $\phi_r=13.87$ and $d=0.086$. The vertical thin line marks the CMB scale $k_*=0.05$ Mpc$^{-1}$.}
\label{fig:PsSR}
\end{figure}

Our resulting fiducial model has the parameter values $M=1.303\times10^{-4}$, $m=2.58\times10^{-6}$, $\phi_r=13.87$, which are the choices in Figs.~\ref{fig:A}--\ref{fig:cs}, and $d=0.086$, which satisfies the observational constraints on $n_s$ and $\alpha_s$. Comparing to Figs.~\ref{fig:A} and \ref{fig:dN}, we see that this model has a relatively fast transition and a moderate violation of slow-roll at the transition. For this set of parameters, we show the resultant scalar power spectrum in Fig.~\ref{fig:PsSR} as computed  by solving numerically the Mukhanov-Sasaki equation~\eqref{eq:muksa} and compare that to the SR formula in Eqs.~\eqref{eq:srpsfor} using the numerical relationship for $\phi(N)$ with $k c_s/aH=1$ (upper panel). The discrepancy, which is quantified as the fractional difference between the solutions and shown in the lower panel, is similar to the error in $\epsilon_H$, as shown in Fig.~\ref{fig:dN}, in that they both peak near the transition where $\A\approx1$. On the other hand, the slow-roll approximation captures the qualitative behavior of the power spectra and errs mainly in causing a shift in the scale $k$ at which the transition occurs. We shall see in the next section that the optimized evaluation of slow-roll parameters can restore accuracy in the CMB regime by correctly fixing this shift.

We can now see how introducing a more rapid transition from G-inflation can solve the observational problem of having too large $r$ for the observed $n_s$.  Namely, the transition mediates a suppression of the power spectrum or a larger red tilt $1-n_s$ than predicted by the slow-roll formula in \S \ref{sec:mod}.   

\begin{figure}[t]
\centering
\includegraphics[keepaspectratio, width=12cm]{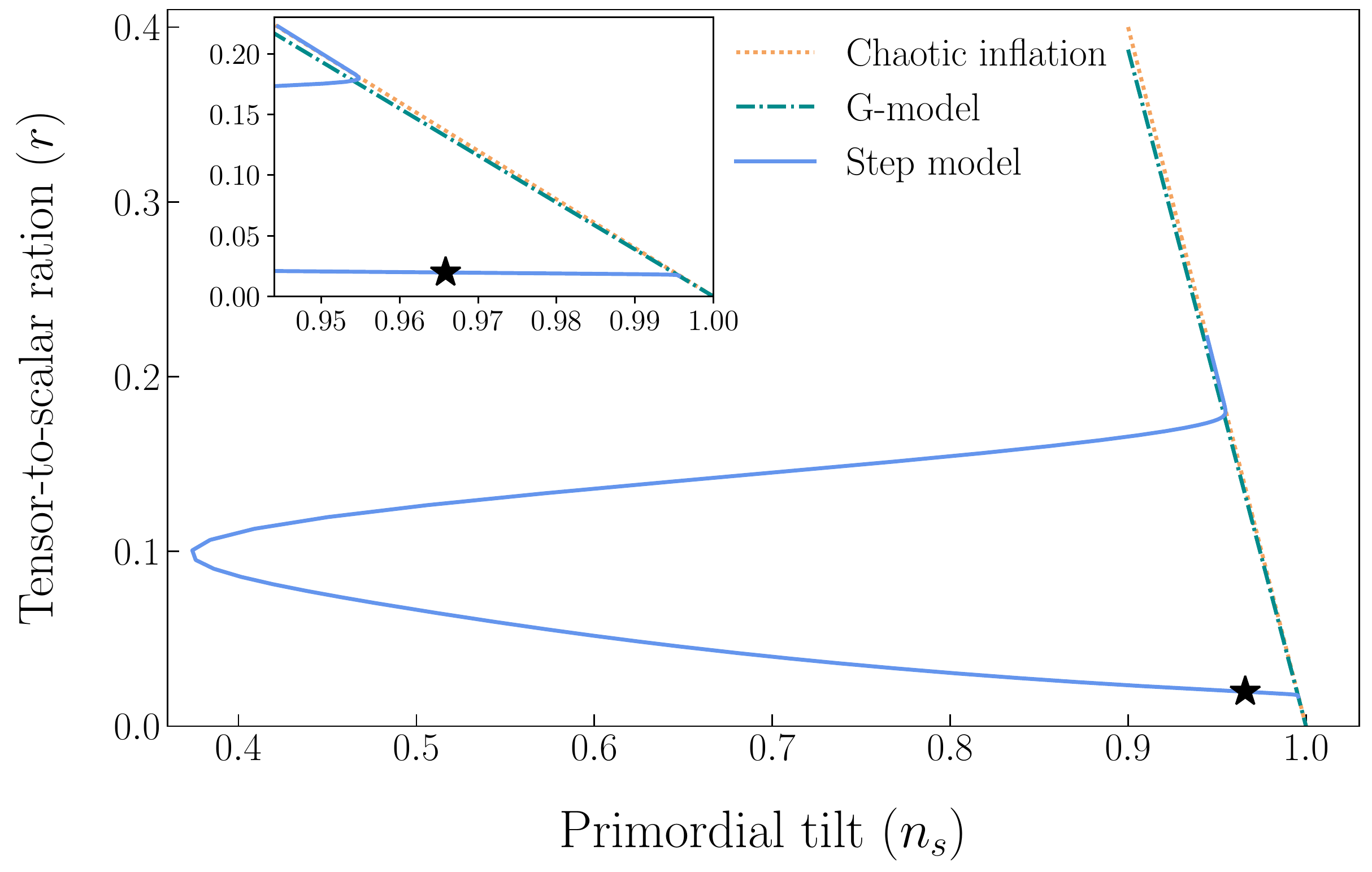}
\caption{$(n_s,r)$ plane for the three models studied here: chaotic inflation~\eqref{eq:rnslimC}, G-model~\eqref{eq:rnslimG} and the Step model with the values $M=1.303\times10^{-4}$, $m=2.58\times10^{-6}$, $\phi_r=13.87$ and $d=0.086$. The wavenumber $k$ varies along the curve in which case the star marks the CMB scale $k_*=0.05$ Mpc$^{-1}$.}
\label{fig:nsrasN}
\end{figure}

In Fig.~\ref{fig:nsrasN} we show the parametric relationship between $r$ and the $n_s$ for same model. The step model starts at the lower right on the G-model curve but deviates sharply to a lower tilt at the transition before returning to the chaotic inflation curve. In this way, the step solves the observational problem of having a low $r$ and a relatively large red tilt $n_s<1$. Note that in Fig.~\ref{fig:nsrasN} the wavenumber $k$ varies along the curve and so only represents the CMB pivot scale at a single point  represented by the star.   This model satisfies observational bounds on $r$ and $n_s$, unlike the G-model and chaotic inflation.
  
\begin{figure}[t]
\centering
\includegraphics[keepaspectratio, width=12cm]{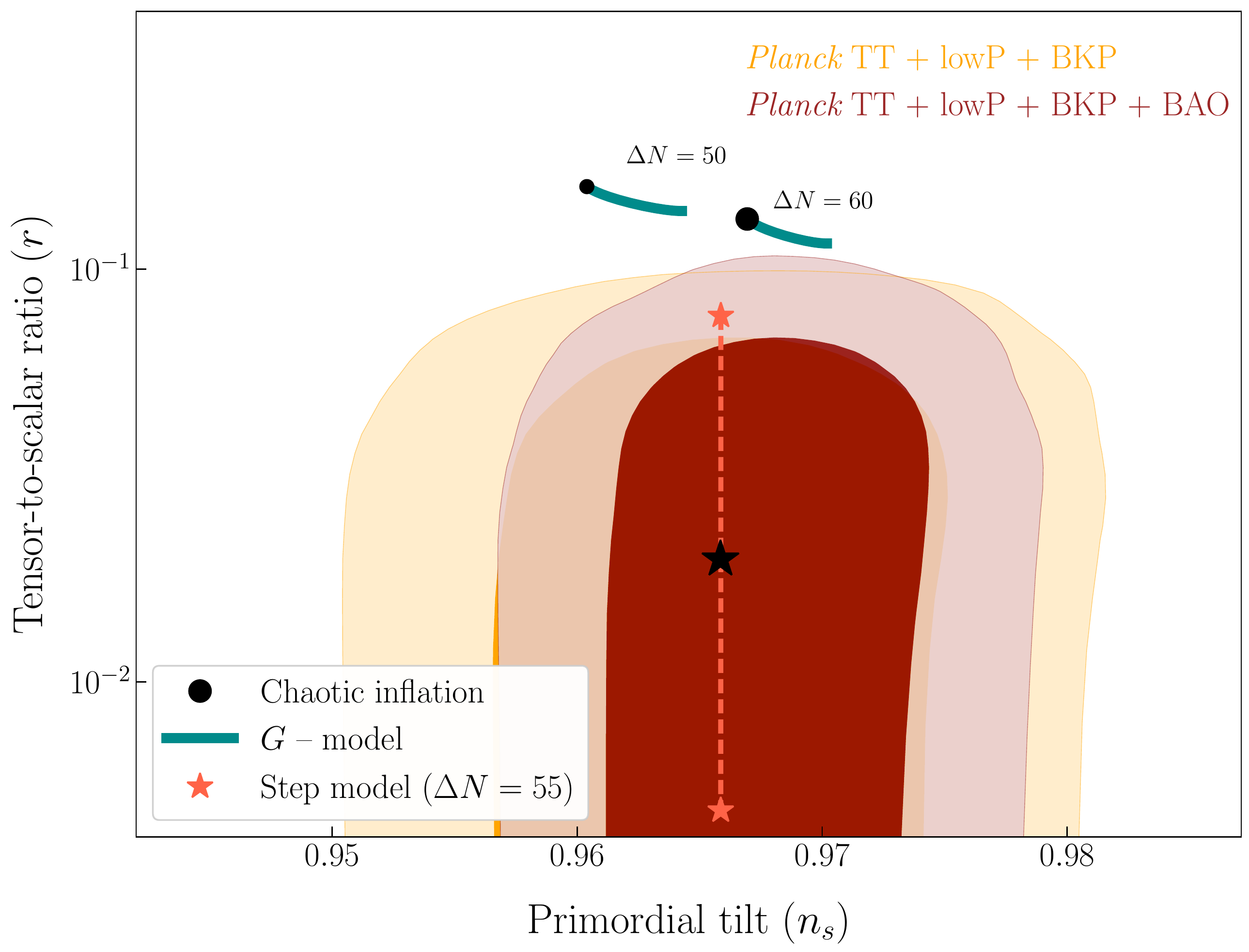}
\caption{68\% and 95\% CL allowed contours from {\it Planck} TT + lowP+ BKP and from {\it Planck} TT + lowP + BKP + BAO in the $(n_s,r)$ plane along with the predictions for the canonical chaotic inflation model, the G-model (both for $\Delta N=50$ and $\Delta N=60$) and the Step model ($\Delta N=55$). In both noncanonical cases, $M$ is let to vary from $10^{-4}$ to $5\times10^{-3}$, where a smaller $M$ value is associated with a smaller value in $r$. The black star marks the fiducial-model value $M=1.303\times10^{-4}$, whereas $M=6.8\times10^{-4}$ ($M=5\times10^{-5}$) for the upper (lower) orange stars endpoints
determined by requiring the scalar running $\alpha_s \approx -0.01$.   Other Step model parameters  are fixed by measurements of $A_s$ and $n_s$ as described in the text.}
\label{fig:nsrplanck}
\end{figure}

Figure~\ref{fig:nsrplanck} depicts the same $(n_s,r)$ plane but now for the fixed pivot scale $k_*$. For the G-model and chaotic inflation we show the mapping at $\Delta N=50,\,60$ to provide a reasonable range of possibilities as in Ref.~\cite{Ohashi:2012wf}, whereas for the Step model we keep $\Delta N=55$. The Galileon mass scale $M$ varies across the curves, where the black star marks the fiducial model $M=1.303\times10^{-4}$, and superimposed are the constraints from the 2015 release of the {\it Planck} collaboration~\cite{Ade:2015lrj}: we separately consider the full temperature auto-correlation spectrum at all multipoles with the polarization spectra at low multipoles only ({\it Planck} TT+lowP) plus the joint results of the Bicep2/Keck and {\it Planck} collaborations (BKP); as well as the {\it Planck} TT+lowP + BKP combination with Baryon Acoustic Oscillation (BAO) measurements.

As previously discussed, one can see that while the canonical chaotic and G- models are in tension with the latest measurements, the Step model allows for a parameter space of values for $M$ which are in good agreement with the data. Following the methodology explained above, for a given value of $M$, the inflaton mass $m$ is fixed to obtain the correct scalar amplitude, while the step parameters $\phi_r$ and $d$ are chosen to keep $n_s$ and $\alpha_s$ fixed. Here we have chosen $\alpha_s \approx -0.01$. Making $M$ smaller allows the Step model to lower the value of $r$ while the transition keeps the CMB scales sufficiently red-tilted.

Furthermore, by varying $M$ away from the fiducial value we encounter two endpoints. As  $M$, and hence $r$, decreases, the increasing value of $\A(0)$ combined with the requirement that $\A<1$ at the end of inflation, places a lower limit on $|\alpha_s|$ for a given $n_s$.   This lower limit exceeds $|\alpha_s|=0.01$ at $r\approx 0.005$ explaining the lower endpoint in Fig.~\ref{fig:nsrplanck}. On the other hand, for large $M$, {CMB scales} are no longer in a fully G-inflationary phase so that $\phi_r$ and $d$ can also no longer be adjusted to match $n_s$ and, more importantly, $\alpha_s \approx -0.01$.

As one might expect, taking a smaller value of $|\alpha_s|$, which still satisfies the Planck constraint, enables a less restrictive upper endpoint that eventually joins with the chaotic or G-model curves. A smaller $|\alpha_s|$ also implies a wider step and   increases the lower limit on $r$  from requiring the transition complete before the end of inflation. A larger $|\alpha_s|$ would have the opposite effects but would begin to be in tension with {\it Planck} constraints.  We thus conclude that for the Step model $r \gtrsim 0.005$, and at the lowest $r$-value  $|\alpha_s|>0.01$, so that tensors and potentially scalar running should be observable with next generation surveys. We comment further on the latter in \S \ref{sec:osr}.

For these observationally viable cases, perturbations on CMB scales were frozen in at the very beginning of the transition.  As we shall see next, this implies that CMB observables can be accurately predicted by the OSR 
approximation which takes into account the variation of the slow-roll parameters.

\section{Generalized slow-roll}
\label{sec:gsr}

In the previous section we have seen that, by introducing a rapid transition from G-inflation to canonical inflation that completes shortly after the CMB scales leave the horizon, we can avoid the observational problems associated with the original G-model. At the transition, the breakdown of the slow-roll approximation requires numerical solutions for full accuracy, especially for large and sharp steps. On the other hand, CMB scales in observationally viable cases are associated with the very beginning of the transition where there is a much milder breakdown of slow-roll. For CMB observables it is therefore possible to develop a better version of slow-roll that is analytic or semi-analytic. This also helps  clarify the phenomenology of the Step model and assists in parameter estimation from the observational data.

Techniques to handle such cases have already been developed for the effective field theory (EFT) of inflation  \cite{Motohashi:2015hpa, Motohashi:2017gqb}, including the Horndeski theory to which our Step model belongs: first the GSR formalism \cite{Stewart:2001cd, Kadota:2005hv, Dvorkin:2009ne, Hu:2011vr, Hu:2014hoa, Motohashi:2015hpa, Miranda:2015cea} allows for formally solving the Mukhanov-Sasaki equation \eqref{eq:muksa}, in which only the size, but not the evolution, of the slow-roll parameters is required to be small.  When the evolution is also slower than the $e$-folding scale, GSR itself can be systematically expanded in the OSR approximation which fixes the evaluation point of the slow-roll parameters to obtain fully analytic solutions.  Since this is the case for the Step model at the beginning of the transition, the OSR approximation is accurate for this model at CMB scales.

\begin{figure}[t]
\centering
\includegraphics[keepaspectratio, width=12cm]{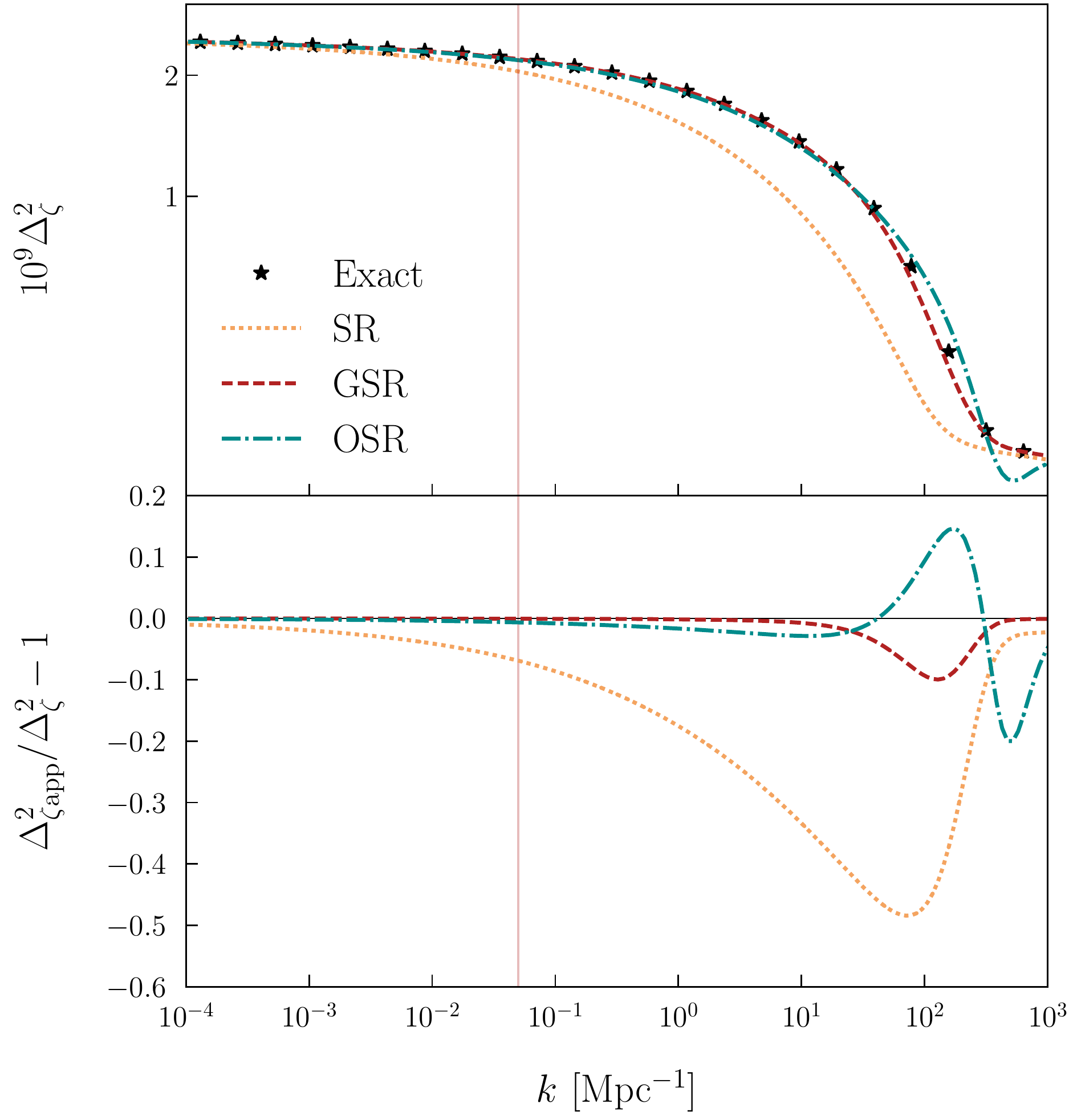}
\caption{Approximations for the scalar  power spectrum of the Step model compared with the exact numerical solution  as in Fig.~\ref{fig:PsSR}: SR \eqref{eq:srpsfor}, GSR \eqref{eq:gsrfor}, and OSR  \eqref{osrscalars}.  The thin red vertical line marks the CMB pivot scale $k_*=0.05$ Mpc$^{-1}$ where GSR and OSR provide a highly accurate description.}
\label{fig:fraccdiff}
\end{figure}

\subsection{GSR}

We can rewrite the Mukhanov-Sasaki equation \eqref{eq:muksa}, as
\be  \label{eq:msgsr}
\difs{y}{x}+\mk{1-\f{2}{x^2}}y=\f{f_{,\chi\chi}-3f_{,\chi}}{f}\f{y}{x^2}~,
\ee
by defining the new variables $y=\sqrt{2c_{i}k}u_i$,  $x \equiv ks_{i}$, $\chi\equiv \ln x$,  for $i=s,t$ with separate source functions
\begin{align}
f_s&\equiv2\pi z_s\sqrt{c_s}s_s=\sqrt{8\pi^2\f{b_s\epsilon_Hc_s}{H^2}}\f{aHs_s}{c_s}~,\notag\\
f_t&\equiv2\pi z_t\sqrt{c_t}s_t=\sqrt{2\pi^2\f{b_tc_t}{H^2}}\f{aHs_t}{c_t}~,
\end{align}
and sound horizons
\begin{equation}
s_{s,t}(N) \equiv \int_{N}^{N_f} \frac{ c_{s,t}}{ a H} \dif{N}  ~,
\end{equation}
for scalars and tensors, respectively.

Notice that the left-hand side of Eq.~(\ref{eq:msgsr}) corresponds to the evolution of the modefunctions in a de Sitter background and thus the right-hand side encodes deviations from the de Sitter solution into the function $f$.   So far we have not made any assumption for the evolution of $\epsilon_H$ or the other slow-roll parameters. In these variables the power spectra, Eqs.~\eqref{eq:standardps}, are given by
\be
\Delta^2_{\zeta,\gamma}=\lim_{x\to0}\bigg{\rvert}\mk{\f{xy}{f}}_{s,t}\bigg{\rvert}^2~.
\ee

If deviations from de Sitter remain small in amplitude, Eq.~(\ref{eq:msgsr}) can  be solved iteratively using Green function methods. Starting with the de Sitter solution of the left-hand side of Eq.~(\ref{eq:msgsr}), {\it i.e.} the Bunch-Davies vacuum, 
\be
y_0(x)=\mk{1+\frac ix}e^{ix}~,
\ee
we can take $y \rightarrow y_0$ on the right-hand side of  Eq.~(\ref{eq:msgsr}) to obtain the
first-order iterative solution (see, {\it e.g.}, \cite{Dvorkin:2009ne} for details), 
\be \label{eq:gsrfor}
\logn\Delta^{2\text{\,(GSR)}}\simeq-\int_0^\infty\f{\dif{x}}{x}W_{,\chi}(x)G\mk{\chi}~,
\ee
where $W(x)$ is a window function given by
\be
W(x)=\frac{3\sin(2x)}{2x^3}-\frac{3\cos(2x)}{x^2}-\frac{3\sin(2x)}{2x}~,
\ee
and $G(\chi)$ is a source function that now encodes all the deviations from the de Sitter solution and it is written as 
\be \label{eq:gsource}
G\equiv-2 \ \logn f+\frac23\mk{\logn f}_{,\chi}~.
\ee

The GSR formula, Eq.~\eqref{eq:gsrfor}, still requires numerical integration, though it remains more computationally efficient than solving Eq.~(\ref{eq:muksa}). Moreover, the source function $G$ provides a model-independent means to connect observational constraints with any inflationary model in the EFT class
\cite{Dvorkin:2011ui,Obied:2017tpd}. The scalar tilt $n_s$ and higher order running coefficients can also be efficiently computed numerically by taking  derivatives of Eq.~\eqref{eq:gsrfor} with respect to the scale $k$.

In Fig.~\ref{fig:fraccdiff}, we compare the GSR approximation to the numerical solution for the same model as in Fig.~\ref{fig:PsSR}. GSR provides accurate predictions for the scalar power spectrum along all values of $k$ and only deviates slightly at the transition due to its large amplitude, which can be improved if desired by iterating to higher order.  At CMB sales of $k_*=0.05$ Mpc$^{-1}$, the approximation is accurate at the $\sim 0.01\%$ level whereas SR deviations are at 7\% level.

\subsection{OSR}
\label{sec:osr}

At CMB scales, the source function in Eq.~(\ref{eq:gsource}) evolves only over timescales greater than an $e$-fold ($\Delta N>1$) as shown  for $\epsilon_H$ in Fig.~\ref{fig:dN}. In this case we can Taylor expand the GSR formula, Eq.~(\ref{eq:gsrfor}), around  a given evaluation epoch to write down approximate analytical formulas for the power spectra, their tilts and runnings. For the traditional slow-roll expansion, the evaluation epoch is chosen as the horizon exit epoch. However, we can optimize it to minimize an error associated with truncation of the Taylor expansion (see \cite{Motohashi:2015hpa, Motohashi:2017gqb} for details). We can then construct the hierarchy of running of power spectrum parameters out of slow-roll parameters associated with the functions $H$, $b_{s,t}$ and $c_{s,t}$. The OSR formulas which take into account a general background given by Eq.~(\ref{eq:Horn}) can then be written to leading order as~\cite{Motohashi:2015hpa, Motohashi:2017gqb}
\begin{align} \label{osrscalars}
\logn\Delta_\zeta^{2\text{\,(OSR)}}&\simeq\logn\mk{\f{H^2}{8\pi^2 b_sc_s\epsilon_H}}-\f{10}{3}\epsilon_H-\f{2}{3}\delta_1-\f{7}{3}\sigma_{s1}-\f{1}{3}\xi_{s1}\Big|_{x=x_1}~,\notag\\
n_s^{\text{\,(OSR)}}-1&\simeq-4\epsilon_H-2\delta_1-\sigma_{s1}-\xi_{s1}-\f{2}{3}\delta_2-\f{7}{3}\sigma_{s2}-\f{1}{3}\xi_{s2}\Big|_{x=x_1}~, \\
\alpha^{\text{\,(OSR)}}_s&\simeq-2\delta_2-\sigma_{s2}-\xi_{s2}-\f{2}{3}\delta_3-\f{7}{3}\sigma_{s3}-\f{1}{3}\xi_{s3}-8\epsilon_H^2-10\epsilon_H\delta_1+2\delta_1^2\Big|_{x=x_1}~\notag,
\end{align}
for scalar, and
\begin{align} \label{osrtensors}
\logn\Delta_{\gamma}^{2\text{\,(OSR)}}&\simeq\logn\mk{\f{H^2}{2\pi^2b_tc_t}}-\f{8}{3}\epsilon_H-\f{7}{3}\sigma_{t1}-\f{1}{3}\xi_{t1}\Big|_{x=x_1}~,\notag\\ 
n_t^{\text{\,(OSR)}}&\simeq-2\epsilon_H-\sigma_{t1}-\xi_{t1}-\f{7}{3}\sigma_{t2}-\f{1}{3}\xi_{t2}\Big|_{x=x_1}~, \\
\alpha_t^{\text{\,(OSR)}}&\simeq-\sigma_{t2}-\xi_{t2}-\f{7}{3}\sigma_{t3}-\f{1}{3}\xi_{t3}-4\epsilon_H^2-4\epsilon_H\delta_1\Big|_{x=x_1}~\notag,
\end{align}
for tensor perturbations. Here $\logn x=\logn x_1\approx1.06$ is the optimized evaluation point, $\alpha_i = d n_i/d\ln k$ is the running of the tilt, and the slow-roll parameters are defined as:
\allowdisplaybreaks
\begin{eqnarray} \label{osrsrparam}
\delta_1&\equiv&\frac12\diff{\,\logn\epsilon_H}{N}-\epsilon_H~,\,\,\,\,\,\,\,\,\,
\delta_{p+1}\equiv\diff{\delta_p}{N}+\delta_p\mk{\delta_1-p\epsilon_H}~,\notag\\ 
\sigma_{i,1}&\equiv&\diff{\, \logn c_i}{N}~,\,\,\,\,\,\,\,\,\,\,\,\,\,\,\,\,\,\,\,\,\,\,\,\,\,\,\,
\sigma_{i,p+1}\equiv\diff{\sigma_{i,p}}{N}~,\\
\xi_{i,1}&\equiv&\diff{\logn b_i}{N}~,\,\,\,\,\,\,\,\,\,\,\,\,\,\,\,\,\,\,\,\,\,\,\,\,\,\,\,\,\,
\xi_{i,p+1}\equiv\diff{\xi_{i,p}}{N}~\notag,
\end{eqnarray}
where $i=s,t$ and $p\geq1$

Finally, the tensor-to-scalar ratio can be computed in the standard way through Eq.~\eqref{eq:rns}. Note however that the ratio is taken at fixed $k$ which in general gives the $x=x_1$ evaluation point at two different $N$ for scalars and tensors, in which case the sound speeds $c_s$ and $c_t$ differ. Figure~\ref{fig:fraccdiff} shows that although the OSR solution  for the scalar power spectrum is slightly less accurate than GSR, it is still a very good approximation with only $\sim 0.6\%$ level deviations at   $k_*=0.05$ Mpc$^{-1}$ (marked by the thin red line).

\begin{figure}[t]
\centering
\includegraphics[keepaspectratio, width=12cm]{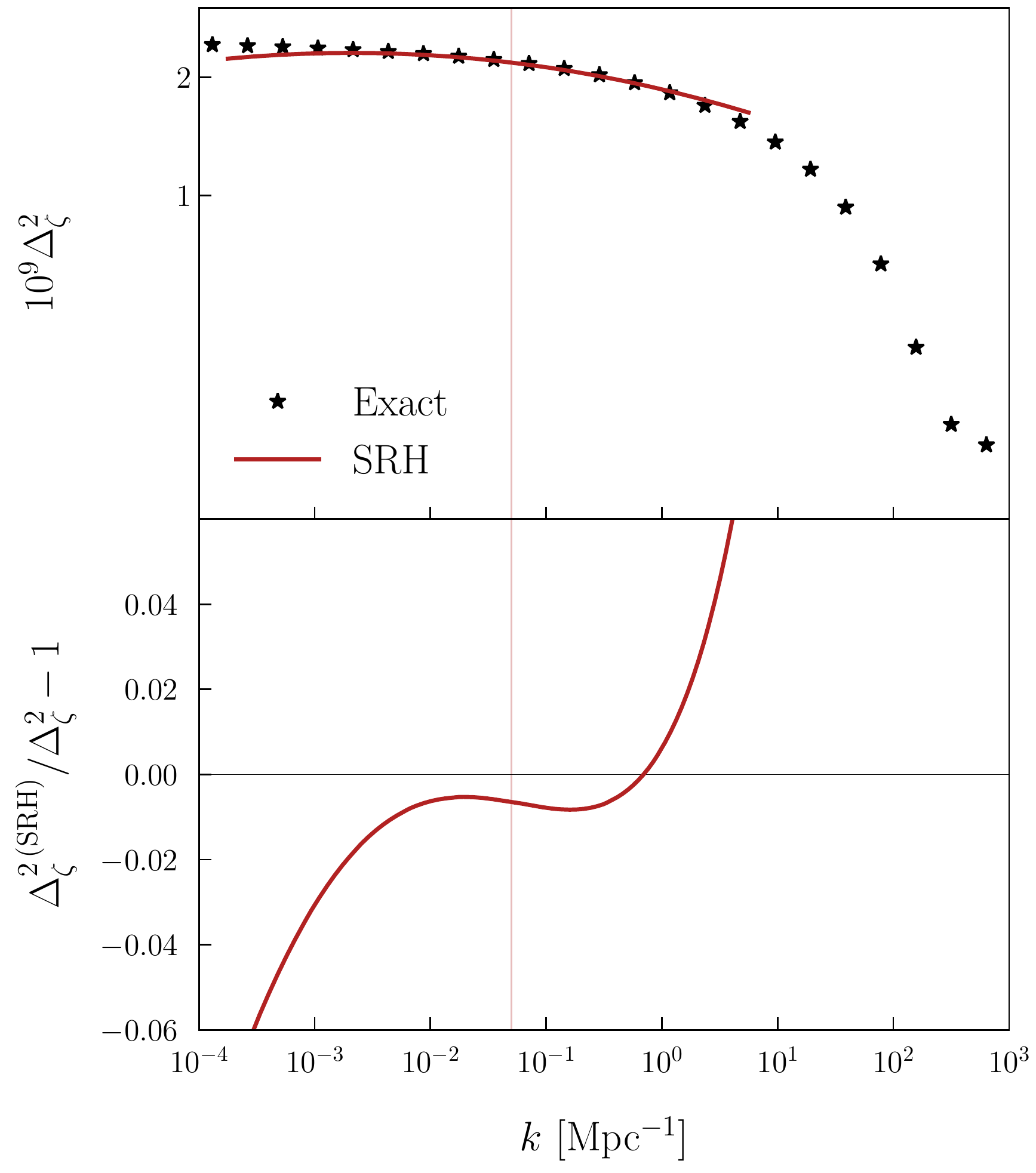}
\caption{Slow-roll hierarchy (SRH) parameterization, Eq.\eqref{eq:SRH}, of the scalar power spectrum with amplitude $A_s$, tilt $n_s$ and running of the tilt $\alpha_s$ evaluated at $k_*$ (thin red line) using OSR  compared with the exact solution as in Fig.~\ref{fig:PsSR}.
The three hierarchy parameters provide a good description for more than two decades around $k_*$.}
\label{fig:fraccdiffSRH}
\end{figure}

Furthermore the hierarchy of OSR coefficients $A_s \equiv \Delta_\zeta^{2(\rm OSR)}(k_*)$, $n_s(k_*)$ and $\alpha_s(k_*)$ define a  local characterization of the scalar power spectrum in the usual way:
\begin{eqnarray} \label{eq:SRH}
\Delta_\zeta^{2(\rm SRH)}(k)  =  A_s \left(\frac{k}{k_*}\right)^{n_s-1 +\frac{1}{2}\alpha_s\ln({k}/{k_*})}~.
\end{eqnarray}
In Fig.~\ref{fig:fraccdiffSRH}, we show that for the decade below or above the pivot scale $k_*$, this three-parameter approximation works extremely well with errors less than $1\%$. This means that observational data in this regime can be analyzed with the usual hierarchy parameterization so long as the implications for inflationary model are extracted from the OSR relations. For example, in the fiducial Step model,  $\alpha_s^{\text{(OSR})}(k_*)=-0.011$ can be compared with the {\it Planck} temperature power spectrum constraint of $\alpha_s = -0.0084 \pm 0.0082$ \cite{Ade:2015lrj}. Unlike the traditional expansion of the SR approximation to second order in parameters, OSR can accurately relate inflationary models to the SRH observables in such cases when $|\alpha_s|$ is of order $|n_s-1|$ \cite{Motohashi:2015hpa}.

Finally as discussed in \S\ref{sec:obs},  the step model allows for a  possible range of values of the running $\alpha_s$ for a given value of $r$. For $|\alpha_s|$ to be small, the transition must be wide, and enforcing that the transition completes before the end of inflation places a lower bound on $|\alpha_s|$. For instance, for $r=0.02$, this corresponds to the constraint $|\alpha_s| \gtrsim 0.002$. Furthermore, this lower bound on $|\alpha_s|$ increases as $r$ decreases as the model must transition from an increasingly enhanced scalar power spectrum within the $\sim 55$ $e$-folds to the end of inflation;  at $r=0.005$, $|\alpha_s|\gtrsim 0.01$.

\section{Conclusions}
\label{sec:con}

G-inflation provides the possibility that inflation is driven by simple potentials, like the  mass term of chaotic inflation, but with more complex kinetic interactions, while still satisfying observational constraints on the scalar and tensor power spectra. We show that  this is not possible with just a cubic Galileon interaction with a constant mass scale since the transition from G-inflation to canonical inflation is too slow, leading to a scalar power spectrum that is either too small relative to the tensors or too close to scale invariant. In addition, failure to complete the transition by the end of inflation leads to gradient instabilities during reheating. By introducing a sufficiently rapid step-like transition, we simultaneously solve both the phenomenological and instability problems of potential driven G-inflation. Although we chose the $m^2\phi^2$ model of chaotic inflation,  steeper monomial potentials also suffer from the same problems which can be solved in the same way.

While a fast transition inevitably leads to a breakdown of the traditional slow-roll approximation at the peak of the transition, we show that for phenomenologically viable models, fluctuations on CMB scales freeze out near the beginning of the transition. By comparing exact numerical solutions with the generalized slow-roll approximation and its optimized expansion, we show how to accurately relate the properties of the G-step model, such as the position and width of the step and two mass scales, to the power spectrum observables through the slow-roll
parameters. In particular, across the scales that are currently precisely measured by the CMB and large-scale structure,  the scalar power spectra can still be described by an amplitude $A_s$, tilt $n_s-1$ and running of the tilt $\alpha_s$.

However the negative running of the tilt can be of order of $n_s-1$ itself unlike in the traditional slow-roll approximation and necessitates  the OSR approximation for its evaluation. In fact for a given tensor-to-scalar ratio $r$, there is a lower bound on $|\alpha_s|$ since the transition must complete within the $\sim 55$ $e$-folds to the end of inflation to avoid gradient instabilities. While the required relatively large running of the tilt can satisfy current constraints if $r \gtrsim 0.005$, it is potentially detectable with future high precision measurements and also suppresses smaller scale structure in observable ways.

\acknowledgments

We thank Eiichiro Komatsu for useful and enlightening discussions.
HR and OM would like to thank the Fermilab Theoretical Physics Department and Kavli IPMU for their hospitality.
HR\ and OM\ were supported in part by MINECO Grant SEV-2014-0398,
PROMETEO II/2014/050,
Spanish Grants FPA2014-57816-P and FPA2017-85985-P of the MINECO, and
European Union's Horizon 2020 research and innovation programme under the Marie Sk\l{}odowska-Curie grant agreements No.~690575 and 674896.
HM\ was supported by JSPS KEKENHI Grant No.~JP17H06359.
WH\ was supported by
U.S.~Dept.\ of Energy contract DE-FG02-13ER41958,  
NASA ATP NNX15AK22G and the Simons Foundation.

\bibliographystyle{JHEP}
\bibliography{ref-Gstep}
\vfill
\end{document}